\documentclass[amsmath,amssymb,prb,superscriptaddress,preprint]{revtex4}
\usepackage{graphicx}
\begin{document}
\title{Tuning thermal transport in nanotubes with topological defects}

\author{Jian Wang}
\email{phcwj@hotmail.com} \affiliation{College of Physical Science and Technology, Yangzhou University, Yangzhou 225002, P. R. China  }

\author{Liang Li}
\affiliation{College of Physical Science and Technology, Yangzhou University, Yangzhou 225002, P. R. China  }

\author{Jian-Sheng Wang}
\affiliation{Center for Computational Science and Engineering and Department of Physics, National University of Singapore, Singapore 117542, Republic of Singapore  }

\date{25 July 2011}

\begin{abstract}
Using the atomistic nonequilibrium Green's function, we find that thermal conductance of carbon nanotubes with presence of topological lattice imperfects is remarkably reduced, due to the strong Rayleigh scattering of high-frequency phonons. Phonon transmission across multiple defects behaves as a cascade scattering  based with the random phase approximation.  We elucidate that  phonon scattering by structural defects is related to the spatial fluctuations of local vibrational density of states(LVDOS).  An effective method of tuning thermal transport in low-dimensional systems through the modulation of LVDOS has been proposed. Our findings provide insights into experimentally controlling thermal transport in nanoscale devices.
\end{abstract}
\pacs{66.70.+f, 44.10.+i}

\maketitle

Carbon nanotubes (CNTs) are interesting materials for nanoscale
electronic devices due to their outstanding electronic and thermal
properties.\cite{nanotubebook1} It is also found that CNTs have some
unusual thermal properties, such as good thermal
conductivity,\cite{thermalpro1,mfph} thermal
rectification\cite{thermalrectification} and thermal
waveguide.\cite{waveguide} Tuning thermal transport in low-dimensional
systems like CNTs not only acts as an challenge for conventional
nanoelectronics and energy conversion on a
chip,\cite{thermoelectric,cnt_thermoelectric} but also gives rise to
an emerging field such as phononic devices and thermal logic
units.\cite{thermallogical} Thermal conductivity has been tailored with some methods such as
the disordered layer,\cite{interacecontrol,individualcontrl} the
isotopic mass disorder\cite{massyang,cnbt,disordered,nidisorder,thermalcontrol}
and the pressure.\cite{pressurecontrol} In contrast with the
high-concentrations of isotopic mass
disorder,\cite{massyang,cnbt,disordered,nidisorder,thermalcontrol} the structural
lattice imperfection in CNTs can be well controlled\cite{defects} during the
growth or by the irradiation process.

In this letter we aim to investigate the effects of multiple topological structure
defects on thermal transport in CNTs through the approach of atomistic
nonequilibrium Green's function(NEGF).\cite{transportation,ejph} We find
that thermal conductance of CNTs can be efficiently tuned with only a few topological lattice defects.  We propose an effective method of tuning thermal conductivity in low-dimensional systems through the modulation of local vibrational density of states(LVDOS). Our findings are interesting to the experimental control of thermal transport in nano devices.

We consider a carbon nanotube connected to heat reservoirs with
different temperatures at each tube end. An energy current flows from
high temperature to low temperature across the tube. To eliminate the
interfacial thermal resistance between the tube and the reservoir, the
heat baths are simulated with the same semi-infinitely long tubes.
Under steady state, the heat current $I$ is
described by NEGF.\cite{ejph} Here we concentrate on the behavior of phonon scatterings
induced by lattice imperfections so that nonlinear scatterings are ignored. In fact, nonlinear phonon
interactions can be included through the calculation of the nonlinear self-energy.\cite{ejph} If the elastic scattering is considered, the heat current expression is reduced to the Landauer formula with thermal conductance $G_{th}$ at temperature $\mathrm{T}$ defined as\cite{ejph}
\begin{equation}
\label{thermal_conductance} \mathrm{G}_{th}=\frac{1}{2\pi}\int_{0}^{\infty}{\mathrm{d}\omega\hbar\omega \mathcal{T}[\omega]  \frac{\partial f}{\partial T}},
\end{equation}
where the phonon transmission $\mathcal{T}[\omega]$ at a given frequency $\omega$ can be calculated through the Caroli formula $\mathcal{T}=\mathrm{Tr}(\mathbf{\mathrm{G}}^{r} \mathbf{\mathrm{\Gamma}}_{L} \mathbf{\mathrm{G}}^{a} \mathbf{\mathrm{\Gamma}}_{R} )$. Here the self-energy density $\mathbf{\mathrm{\Gamma}}_{\alpha}=i(\Sigma^{r}_{\alpha}-\Sigma^{a}_{\alpha}), \alpha=\mathrm{L},\mathrm{R}$ for the left/right lead,  $\mathbf{\mathrm{G}}^{r}$ and $\mathbf{\mathrm{G}}^{a}$ are the retarded/advanced Green's function, and $\mathrm{f}$ the Bose distribution for phonons.

To calculate the NEGF atomistically, the dynamic matrix are derived from the second-order derivative of
the second-generation Brenner potential\cite{brenner,revisedbrenner} with respect to displacements
after the structures are optimized with the same potential. The Brenner potential can well reproduce the
phonon dispersion relations for CNTs.\cite{revisedbrenner}  In comparison with the first-principle density-functional theory approach, this
empirical potential is short-ranged and more tractable for systems with large number of atoms.\cite{revisedbrenner}  During our
simulations, there are thousands of atoms, which are beyond the ability of any first-principle method.

\begin{figure}[hb]
\includegraphics[width=0.65\columnwidth]{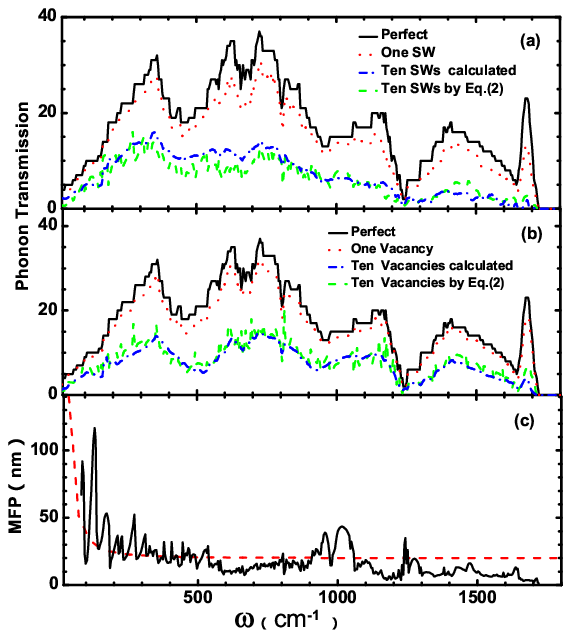}
\caption{\label{fig:transmission} (Color online).  Figures $\bf(a)$ and $\bf(b)$. Phonon transmission
as a function of frequency for the defective CNT $(10,10)$.  The  green dashed lines in figures  are derived from the cascade scattering model. Figure$\bf(c)$.  The frequency dependence of the length of phonon mean free path(MFP) for the CNT with one Stone-Wales defect. The red dashed line illustrates the quadratic relation ($\propto\omega^{-2}$) for a guide view.  }
\end{figure}

The calculated phonon transmissions for the CNT $(10,10)$  through
the NEGF are shown in  Fig.~\ref{fig:transmission}(a) and Fig.~\ref{fig:transmission}(b). With the increase of the number of
structural defects, it can be seen from the figure that phonon transmission for high-frequency decrease rapidly,
irrespective of Stone-Wales or  vacancy defects.  To understand this behavior, we first investigate the phonon length of
mean-free path(MFP). For perfect CNTs, the length of phonon MFP is estimated\cite{mfph}
to be in the order of magnitude of micrometers at room temperature. Here the length of CNTs considered  is a
few nanometers. Thus, the transport process across the defect
scattering can be described as ballistic-diffusive,\cite{mfph,cnbt,ejph} where the total phonon transmission at a given frequency $\omega$ can be written\cite{mfph,cnbt,transportation} as
$\mathcal{T}[\omega]=\mathcal{T}_{p}[\omega]l_0/(L+l_0)$. Here $\mathcal{T}_{p}$ is the phonon transmission for the perfect tube at frequency $\omega$, $l_0$ the length of phonon MFP, and $L$ the length of CNT.  This formula has
been used for estimating the length of phonon MFP.\cite{cnbt} The calculated length of phonon
MFP is shown in Fig.~\ref{fig:transmission}(c), where we can find that
the length of  MFP decreases with the frequency in the quadratic form
($\propto\omega^{-2}$).  For $d$ dimensional systems, the Rayleigh scattering
theory\cite{disordered}  gives the length of the MFP as a function of
$\propto\omega^{-(d+1)}$. The CNTs can be assumed as quasi-one-dimensional systems with $d=1$ such that the phonon MFP has
a quadratic dependence($\propto\omega^{-2}$) by the theory of Rayleigh scattering. The frequency dependency of the phonon MFP in Fig.~\ref{fig:transmission}(c) agrees well with the quadratic relation so that the phonon scattering by defects in CNTs can be understood in terms of Rayleigh
scattering. We thus can conclude that the larger decrease of phonon transmission in the high-frequency range results from strong scatterings of high-frequency phonons, due to the rapid decay ($\propto\omega^{-2}$) of MFP characterized by Rayleigh scattering.

Next we consider the behavior of phonon transmission across  multiple topological lattice defects.  In the
ballistic-diffusive transport regime, the phonon transmission
$\mathcal{T}_{N}$ across $N$ defects can be described by the
cascade scattering model \cite{transportation,cnbt}
\begin{equation}
\label{multiple_scattering} \frac{1}{\mathcal{T}_{N}} =\frac{N}{\mathcal{T}_{1}} - \frac{N-1}{\mathcal{T}_{p}},
\end{equation}
where $\mathcal{T}_{1}$ is the transmission across single defect, $N$ the number of defects, and
$\mathcal{T}_{p}$ the  transmission for the perfect tube. It is straightforward that the quantity
$({\mathcal{T}_{p}}-{\mathcal{T}_{1}})/\mathcal{T}_{1}$ in
Eq.~(\ref{multiple_scattering}) has the additive property with the
transmission probability $\mathcal{T}_{1}$ for each scatter placed in
cascade.\cite{transportation}  The calculated
phonon transmission from the cascade model is shown in Fig.~\ref{fig:transmission}(a) and  Fig.~\ref{fig:transmission}(b) denoted by the green dashed lines. We can find that phonon transmissions derived from  Eq.~(\ref{multiple_scattering}) agree well with the atomistic calculations by NEGF for both Stone-Wales and vacancy defects.  After performing computations for longer CNTs with more structural defects, we have also observed such consistency between the direct numerical calculations by NEGF and the predictions of the cascade scattering model. The reason for the good validity of the cascade model can be explained as follows.  Note that Eq.~(\ref{multiple_scattering}) is associated with a random phase approximation for wave interferences.\cite{transportation} Unlike the point isotope-disorder, structural defects introduce a large area of lattice distortion such that it is difficult to keep the phonon waves coherent. Therefore, the random phase serves as a good approximation for the scattered waves. When the concentration of structural defects is not too high, we think that the cascade scattering model  can well describe phonon scatterings over multiple structural defects.

\begin{figure}[hb]
\includegraphics[width=0.75 \columnwidth]{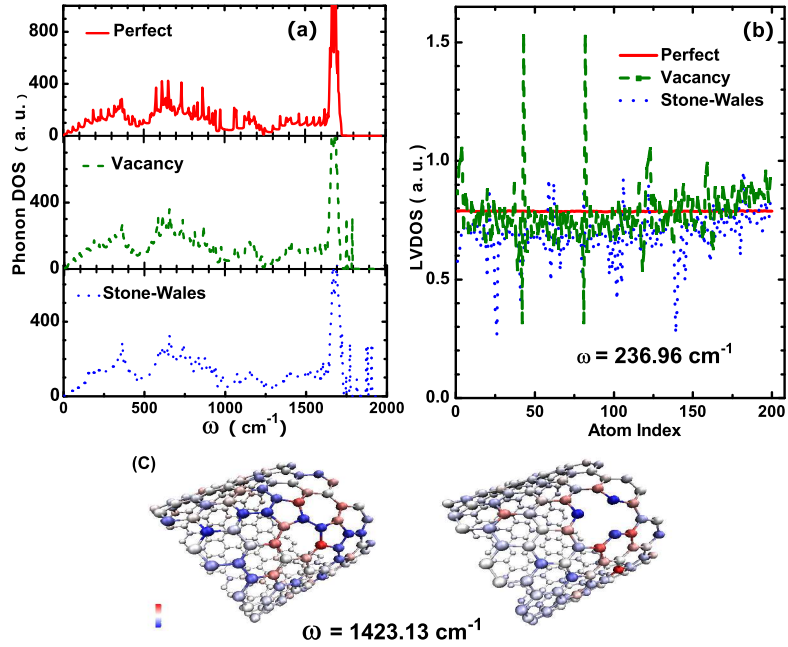}
\caption{\label{fig:gdos} (Color online).  Figure$\bf{(a)}$. The phonon
density of states(DOS) as a function of frequency.  Figure$\bf{(b)}$. A
profile of local vibrational density of states(LVDOS) at frequency
$\omega=236.96 cm^{-1}$.    Figure$\bf{(c)}$. Atomistic view of LVDOS for the CNT$(10,10)$ with a SW defect and a vacancy defect at frequency $\omega=1423.12 cm^{-1}$, respectively. }
\end{figure}

Further, we elucidate the mechanisms of the scattering of phonon by structural defects through calculating the phonon density of
states(DOS) using NEGF.  The phonon DOS and LDOS can be expressed as the imaginary part of the Green's
function as $\mathrm {DOS}[\omega]=-2\omega\mathrm{\bf Im}\mathrm{Tr}(\mathrm{G}^{r})/\pi$ and $\mathrm{LDOS}[\omega]_{ll}=-2\omega\mathrm{\bf
Im}\mathrm{Tr}_{x,y,z}(\mathrm{G}^{r})_{l,l}/\pi$, respectively. The
subscript $l$ is the index of the diagonal term of the Green's function
matrix for each atom. The trace operation $\mathrm{Tr}_{x,y,z}$ for
each atom in the $\mathrm {LDOS}$ is only carried out in three directions.  It can be seen from Fig.~\ref{fig:gdos}(a) that
there is an appearance of some extended spectra of vibrational states
for the defective tubes at $\omega > 1748cm^{-1}$. These
extended spectra correspond to the localized vibrations of the defect
lattices. Fig.~\ref{fig:gdos}(b) shows a profile of the LDOS. We can find that the perfect CNT has a uniform distribution of LDOS, while the LDOS for the defective tubes fluctuates with the positions of atoms, especially for atoms located in defects. The different phonon modes are scattered by the fluctuation of LDOS as shown in Fig.~\ref{fig:gdos}(b). An atomistic view of LDOS  is further demonstrated in Fig.~\ref{fig:gdos}(c).  The LDOS provides a direct view of spatial fluctuations of vibrational states around the defective tubes. The local density of states for electrons can be measured by scanning tunneling microscopy, but the experimental measurement of the phonon LDOS has not been realized. We think that the measurement of local vibrational density of states is not only important to understanding thermal transport in nanoscale systems but also is vital to developing nano devices.

\begin{figure}[hb]
\includegraphics[width=0.65\columnwidth]{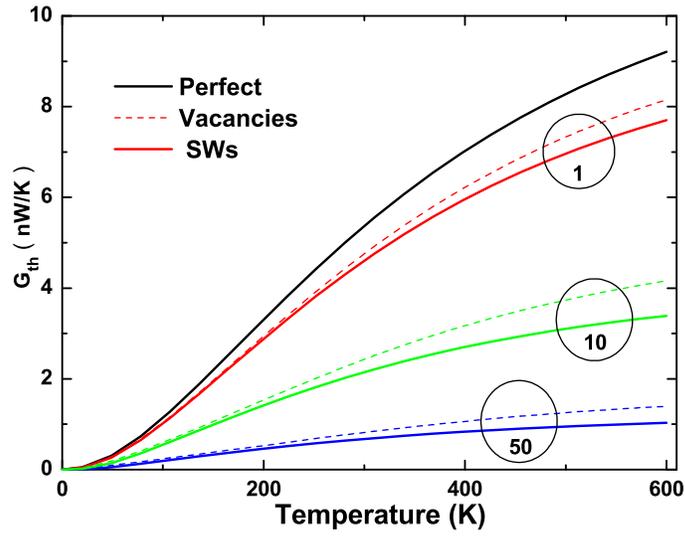}
\caption{\label{fig:conductance} (Color online). The temperature
dependence of thermal conductance for the perfect and defective CNTs. The number of defects are indicated by the number in the circle. The color dashed lines represent thermal conductance for the CNTs with vacancies while the color solid lines are for the SW defective tubes.  }
\end{figure}

Finally, thermal conductance as a function of temperature is shown in Fig.~\ref{fig:conductance}. The solid line on the top represents the ballistic upper bound for the thermal conductance of the perfect CNT.  With the increase of number of defects, we can find that thermal conductance rapidly decreases and varies little with temperature at high temperature.  The reason for this large reduction and small variations is that high-frequency phonons which contribute to thermal transport at high temperature is strongly scattered by the structural defects in terms of Rayleigh scattering.  In contrast to thermal conductance of CNTs with isotope disorders,\cite{massyang,cnbt,disordered,thermalcontrol} we find that even a few structural defects in CNTs can lead to a strong suppression of thermal transport by one order of magnitude.  We further propose that the topological structural defects can offer an effective method of tuning thermal transport in low-dimensional systems like CNTs through the modulation of LDOS.  One advantage of such approach is that topological lattice defects can be well controlled during the growth or by irradiations. Compared with the molecular dynamical simulations,\cite{cntluke,cntAlaghemandi} a decrease of thermal conductance with temperature is not observed at high temperatures due to the neglect of nonlinear phonon-phonon interaction during our calculations.

In summary, using the atomistic NEGF we find that thermal conductance of CNTs with presence of topological imperfects is remarkably reduced. High-frequency phonons are strongly scattered by structural defects, which can be characterized by one-dimensional Rayleigh scattering. In comparison with that of single structural defect in CNTs,\cite{thermalyamoto,mingodef,wangdef} we show that phonon scattering across multiple defects can be well described by the cascade model with the random phase approximation. We also demonstrate that phonon scattering is related to the spatial fluctuations of LDOS. An effective method of tuning thermal transport in low-dimensional systems by modulating LDOS has been proposed. We expect our findings can contribute insights to the experimental control of thermal transport in nanoscale devices.

J.W. acknowledges the support from National Natural Science
Foundation of China (NSFC) under the grant 10705023 and 11075136, as
well as from Jiangsu Natural Science Foundation(BK2009180) . J.-S. W. is supported by a URC grant R-144-000-257-112.

\newpage

\end{document}